\documentclass[12pt]{article}

\newcommand{\xb}{\mbox{\boldmath $x$}}
\newcommand{\pb}{\mbox{\boldmath $p$}}
\newcommand{\ppb}{\mbox{\boldmath $P$}}

\newcommand{\hl}{\hat{L}}
\newcommand{\hp}{\hat{P}}
\newcommand{\tp}{\tilde{p}}
\newcommand{\tpp}{\tilde{P}}
\newcommand{\txx}{\tilde{X}}
\newcommand{\tbeta}{\tilde{\beta}}
\newcommand{\tomega}{\tilde{\omega}}
\def\diag{\mathop{\rm diag}\nolimits}

\title{
Lorentz-covariant deformed algebra with minimal length and application to the ($1+1$)-dimensional
Dirac oscillator}
\author{C Quesne$^1$  and  V M Tkachuk$^2$\\
$^1$ {\small Physique Nucl\'eaire Th\'eorique et Physique
Math\'ematique,  Universit\'e Libre de Bruxelles,} \\ 
{\small Campus de la Plaine CP229, Boulevard~du Triomphe, B-1050 Brussels,
Belgium}\\ 
$^2$ {\small Ivan Franko Lviv National University, Chair of Theoretical
Physics,}\\
{\small 12, Drahomanov Street, Lviv UA-79005, Ukraine}\\
{\small E-mail: cquesne@ulb.ac.be and tkachuk@ktf.franko.lviv.ua}}
\date{ }
\begin{document}
\baselineskip=20pt plus 1pt minus 1pt
\maketitle
\begin{abstract}
The $D$-dimensional $(\beta, \beta')$-two-parameter deformed algebra introduced by Kempf is generalized
to a Lorentz-covariant algebra describing a ($D+1$)-dimensional quantized spacetime. In the $D=3$ and
$\beta=0$ case, the latter reproduces Snyder algebra. The deformed Poincar\'e transformations leaving the
algebra invariant are identified. It is shown that there exists a nonzero minimal uncertainty in position
(minimal length). The Dirac oscillator in a ($1+1$)-dimensional spacetime described by such an algebra is
studied in the case where $\beta'=0$. Extending supersymmetric quantum mechanical and shape-invariance
methods to energy-dependent Hamiltonians provides exact bound-state energies and wavefunctions.
Physically acceptable states exist for $\beta < 1/(m^2 c^2)$. A new interesting outcome is that, in
contrast with the conventional Dirac oscillator, the energy spectrum is bounded.
\end{abstract}

\vspace{0.5cm}


\newpage
\setcounter{footnote}{2}
\section{Introduction}

Many years ago, Snyder~\cite{snyder47a} proposed to abandon the assumption of continuous spacetime. In
his Lorentz-covariant quantized spacetime, the existence of a natural unit of length forced him to drop the
usual hypothesis of commutativity of coordinates. Later on, he studied how to deal with the electomagnetic
field in such a context~\cite{snyder47b}. Then, during several decades, there were only a few works on this
subject~\cite{yang, hellund, fischbach, hamilton}.\par
%
%
Recently, there has been an increasing interest in studying the impact of noncommutativity of coordinates
on the properties of quantum systems. It is motivated by several independent lines of investigation in string
theory and quantum gravity, which suggest the existence of a finite lower bound to the possible resolution
of length (see, e.g., \cite{gross, maggiore, witten}). This would quantum mechanically be described as a
nonzero minimal uncertainty in position, which could be obtained through small quadratic corrections to the
canonical commutation relations~\cite{kempf94, kempf95, hinrichsen, kempf97}.\par
%
%
One of the interesting issues arising in such a framework consists in investigating the influence of the
minimal length assumption on the energy spectrum of quantum systems. Solving quantum mechanical
problems with deformed canonical commutation relations, however, usually turns out to be much more
difficult than with conventional ones, so that only a few cases have been considered in such a context.\par
%
%
An exact solution to the one-dimensional harmonic oscillator problem has been provided by solving the
corresponding Schr\"odinger equation in momentum representation~\cite{kempf95}. This approach has been
extended to $D$ dimensions~\cite{chang}. Some perturbative~\cite{brau, stetsko},
numerical~\cite{benczik} or one-dimensional exact~\cite{fityo} results have been obtained for the hydrogen
atom. The harmonic oscillator system, both in one and $D$ dimensions, has also been considered~\cite{cq03,
cq04} in the framework of supersymmetric quantum mechanics (SUSYQM) by using shape-invariance
techniques (for some reviews see, e.g., \cite{cooper, junker}). Such an approach has been shown~\cite{cq04}
to be especially useful when one assumes nonzero minimal uncertainties in both position and momentum.\par
%
%
The only relativistic problem that has been exactly solved in a deformed space with minimal length is the Dirac
oscillator~\cite{cq05}. This system was introduced in a conventional framework many years ago~\cite{ito}.
The interest in the problem was revived later on and the name `Dirac oscillator' was coined to refer
to it~\cite{moshinsky}. Since then, the Dirac oscillator has aroused much interest both because it provides
one of the few examples of exactly solvable Dirac equation and because it can be applied to a variery of
physical problems (for a list of rerences see~\cite{cq05}).\par
%
%
As a matter of fact, the deformed algebra with minimal length, introduced in~\cite{kempf94, kempf95,
hinrichsen, kempf97} and applied to quantum mechanical problems in~\cite{kempf95, chang, brau, stetsko,
benczik, fityo, cq03, cq04, cq05}, is a nonrelativistic one. This algebra is very similar to Snyder one, but, in
contrast with the latter, it is not Lorentz covariant and therefore violates Lorentz symmetry. In this paper,
we plan to eliminate such an undesirable feature by proposing a straightforward generalization of the
algebra, which is Lorentz covariant and contains Snyder algebra as a special case.\par
%
%
In section~2, after introducing our new algebra, we determine its transformation properties under
(deformed) Poincar\'e algebra and study its predictions regarding minimal uncertainties. In section~3, we
consider one of the simplest relativistic quantum systems, namely the ($1+1$)-dimensional Dirac oscillator,
and we calculate its bound-state spectrum and wavefunctions in the space with minimal length described by
our new Lorentz-covariant deformed algebra. Finally, section~4 contains the conclusion.\par
%
%
\section{Lorentz-covariant deformed algebra}

To start with, let us review the notations to be used in the present paper. In the conventional
($D+1$)-dimensional continuous spacetime, coordinates are denoted by contravariant ($D+1$)-vectors
\begin{equation}
  x^{\mu} = (x^0, x^1, x^2, \ldots, x^D) = (x^0, x^i) = (ct, \xb)
\end{equation}
where Greek (respectively latin) indices run over 0, 1, 2, \ldots, $D$ (respectively 1, 2, \ldots, $D$). The
corresponding covariant ($D+1$)-vectors are given by
\begin{equation}
  x_{\mu} = (x_0, x_1, x_2, \ldots, x_D) = (x_0, x_i) = (ct, - \xb) = g_{\mu\nu} x^{\nu}
\end{equation}
with $g_{\mu\nu} = g^{\mu\nu} = \diag(1, -1, -1, \ldots, -1)$. Contravariant and covariant momenta are
similarly defined as  
\begin{equation}
  p^{\mu} = (p^0, p^1, p^2, \ldots, p^D) = (p^0, p^i) = \left(\frac{E}{c}, \pb\right)
\end{equation}
\begin{equation}
  p_{\mu} = (p_0, p_1, p_2, \ldots, p_D) = (p_0, p_i) = \left(\frac{E}{c}, - \pb\right) = g_{\mu\nu}
  p^{\nu}.
\end{equation}
\par
%
%
In the coordinate representation of quantum mechanics, the momentum operators read
\begin{equation}
  p^{\mu} = {\rm i} \hbar \frac{\partial}{\partial x_{\mu}} = {\rm i} \hbar g^{\mu\nu} \frac{\partial}
  {\partial x^{\nu}} = \left({\rm i} \hbar \frac{\partial}{\partial (ct)}, - {\rm i} \hbar \frac{\partial}{\partial
  x^1}, - {\rm i} \hbar \frac{\partial}{\partial x^2}, \ldots, - {\rm i} \hbar \frac{\partial}{\partial x^D}\right)
\end{equation}
while the position operators are replaced by the corresponding variables. We will instead use here the
momentum representation, wherein the position operators become
\begin{equation}
  x^{\mu} = - {\rm i} \hbar \frac{\partial}{\partial p_{\mu}} = - {\rm i} \hbar g^{\mu\nu} \frac{\partial}
  {\partial p^{\nu}} = \left(- {\rm i} \hbar \frac{\partial}{\partial p^0}, {\rm i} \hbar \frac{\partial}{\partial
  p^1}, {\rm i} \hbar \frac{\partial}{\partial p^2}, \ldots, {\rm i} \hbar \frac{\partial}{\partial p^D}\right)
\end{equation}
whereas the momentum variables are substituted for the related operators. The commutation relations of
these operators are given by
\begin{equation}
  [x^{\mu}, p^{\nu}] = - {\rm i} \hbar g^{\mu\nu} \qquad [x^{\mu}, p_{\nu}] = - {\rm i} \hbar
  \delta^{\mu}_{\nu}.   \label{eq:canonical}
\end{equation}
\par
%
%
\subsection{Generalization of the $D$-dimensional deformed algebra with minimal length to a
Lorentz-covariant one}

The $D$-dimensional deformed algebra considered in~\cite{kempf95, kempf97, chang} is characterized by
modified commutation relations, which, in the above-mentioned notations, read
\begin{eqnarray}
  && [X^i, P^j] = - {\rm i} \hbar [(1 + \beta \ppb^2) g^{ij} - \beta' P^i P^j] \nonumber \\
  && [X^i, X^j] = {\rm i} \hbar \frac{2\beta - \beta' + (2\beta + \beta') \beta \ppb^2}{1 + \beta \ppb^2}
        (P^i X^j - P^j X^i) \nonumber \\
  && [P^i, P^j] = 0  \label{eq:Kempf-com}
\end{eqnarray}
where $\beta$ and $\beta'$ are two very small non-negative deforming parameters. It gives rise to
(isotropic) nonzero minimal uncertainties in the position coordinates $(\Delta X^i)_0 = (\Delta X)_0 = \hbar
\sqrt{D\beta + \beta'}$.\par
%
%
In the momentum representation, the deformed position and momentum operators $X^i$, $P^i$ are
represented by
\begin{equation}
  X^i = (1 + \beta \pb^2) x^i + \beta' p^i (\pb \cdot \xb) + {\rm i} \hbar \gamma p^i \qquad P^i = p^i
  \label{eq:Kempf-op}
\end{equation}
where $x^i = {\rm i} \hbar \partial/\partial p^i$, $p^i$ satisfy equation (\ref{eq:canonical}) and $\gamma$
is an arbitrary real constant, which does not influence the commutation relations (\ref{eq:Kempf-com}). Such
a constant, however, affects the weight function in the scalar product in momentum space
\begin{equation}
  \langle\psi | \phi\rangle = \int \frac{d^D\pb}{[1 + (\beta + \beta') \pb^2]^{\alpha}}\,  \psi^*(\pb)
  \phi(\pb)
\end{equation}
where
\begin{equation}
  \alpha = \frac{2\beta + \beta'(D+1) - 2\gamma}{2(\beta + \beta')}. \label{eq:Kempf-alpha}
\end{equation}
These definitions ensure the Hermiticity of the operators (\ref{eq:Kempf-op}).\par
%
%
To convert the non-Lorentz-covariant algebra (\ref{eq:Kempf-com}) into a Lorentz-covariant one, let us
replace $\pb^2$ and $\pb \cdot \xb$ in (\ref{eq:Kempf-op}) by the Lorentz-invariant expressions $\pb^2 -
(p^0)^2 = - p_{\nu} p^{\nu}$ and $\pb \cdot \xb - p^0 x^0 = - p_{\nu} x^{\nu}$, respectively. Instead of
(\ref{eq:Kempf-op}), we therefore consider the operators
\begin{equation}
  X^{\mu} = (1 - \beta p_{\nu} p^{\nu}) x^{\mu} - \beta' p^{\mu} p_{\nu} x^{\nu} + {\rm i} \hbar
  \gamma p^{\mu} \qquad P^{\mu} = p^{\mu} \label{eq:alg-gen}
\end{equation}
which are ($D+1$)-vectors. A straightforward calculation shows that their commutators are given by
\begin{eqnarray}
  && [X^{\mu}, P^{\nu}] = - {\rm i} \hbar [(1 - \beta P_{\rho} P^{\rho}) g^{\mu\nu} - \beta' P^{\mu} 
        P^{\nu}] \nonumber \\
  && [X^{\mu}, X^{\nu}] = {\rm i} \hbar \frac{2\beta - \beta' - (2\beta + \beta') \beta P_{\rho}
        P^{\rho}}{1 - \beta P_{\rho} P^{\rho}} (P^{\mu} X^{\nu} - P^{\nu} X^{\mu}) \nonumber \\
  && [P^{\mu}, P^{\nu}] = 0.  \label{eq:alg-com}
\end{eqnarray}
We shall assume that, in (\ref{eq:alg-gen}) and (\ref{eq:alg-com}), $\beta$ and $\beta'$ are two very small
non-negative parameters.\par
%
%
As it may be guessed from the ($D+1$)-vector notation and as it will be explicitly checked in section~2.2, the
algebra generated by $X^{\mu}$ and $P^{\mu}$ is Lorentz invariant. It may be considered as a
generalization of Snyder algebra~\cite{snyder47a}, since the latter is recovered for $D=3$ and $\beta =
\gamma = 0$.\footnote{The correspondence between our notations and those of Snyder is $X^0 \to ct$,
$X^1 \to x$, $X^2 \to y$, $X^3 \to z$, $P^0 \to - p_t/c$, $P^1 \to p_x$, $P^2 \to p_y$, $P^3 \to
p_z$, $\beta' \to (a/\hbar)^2$.} It should be stressed that although the nonrelativistic deformed algebra
(\ref{eq:Kempf-com}) can be obtained from (\ref{eq:alg-com}) by dropping the term
$(P^0)^2$ from
$P_{\rho} P^{\rho} = (P^0)^2 - \ppb^2$, this procedure is not equivalent to the nonrelativistic limit $c \to
\infty$. Hence the Lorentz-covariant deformed algebra (\ref{eq:alg-com}) is an entirely new algebra, which
cannot be reduced to the nonrelativisitc deformed one, proposed by Kempf.\par
%
%
It can be easily shown that the new operators $X^{\mu}$ and $P^{\mu}$, defined in (\ref{eq:alg-gen}), are
Hermitian with respect to a modified scalar product in momentum space
\begin{equation}
  \langle\psi | \phi\rangle = \int \frac{d^D\pb}{[1 - (\beta + \beta') p_{\nu} p^{\nu}]^{\alpha}}\, 
  \psi^*(p^{\mu}) \phi(p^{\mu})  \label{eq:sc}
\end{equation}
where $\alpha$ is now defined by
\begin{equation}
  \alpha = \frac{2\beta + \beta'(D+2) - 2\gamma}{2(\beta + \beta')} \label{eq:alpha}
\end{equation}
instead of (\ref{eq:Kempf-alpha}). In particular, for $D=3$ and $\beta=\gamma=0$, equation
(\ref{eq:alpha}) yields $\alpha = 5/2$, in accordance with the corresponding result obtained by
Snyder~\cite{snyder47a}.\par
%
%
One may observe that the weight function in (\ref{eq:sc}) may become singular if we allow the energy $E =
cP^0 = cp^0$ to take very large values. This indicates that states with such energies must be considered as
unphysical. In other words, the energy of physically acceptable states satisfies the condition
\begin{equation}
  (\beta + \beta') (p^0)^2 < 1  \label{eq:cond-accept}
\end{equation}
for which the weight function in (\ref{eq:sc}) is free from singularities.\par
%
%
\subsection{Transformation properties under deformed Poincar\'e algebra}

To start with, let us consider the effect of a standard infinitesimal proper Lorentz transformation
\begin{eqnarray}
  && X^{\prime\mu} = X^{\mu} + \delta X^{\mu} \qquad \delta X^{\mu} = \delta
         \omega^{\mu}_{\hphantom{\mu}\nu} X^{\nu} \nonumber \\
  && P^{\prime\mu} = P^{\mu} + \delta P^{\mu} \qquad \delta P^{\mu} = \delta
         \omega^{\mu}_{\hphantom{\mu}\nu} P^{\nu}  \label{eq:Lorentz-tr}  
\end{eqnarray}
with 
\begin{equation}
  \delta \omega_{\mu\nu} = - \delta \omega_{\nu\mu} \in \mathbf{R}  \label{eq:omega}
\end{equation}
on the deformed algebra (\ref{eq:alg-com}). The first commutation relation becomes
\begin{equation}
  [X^{\prime\mu}, P^{\prime\nu}] = [X^{\mu}, P^{\nu}] + [\delta X^{\mu}, P^{\nu}] + [X^{\mu},
  \delta P^{\nu}]. 
\end{equation}
On inserting (\ref{eq:Lorentz-tr}) on its right-hand side and using (\ref{eq:alg-com}), (\ref{eq:omega}), we
obtain
\begin{equation}
  [X^{\prime\mu}, P^{\prime\nu}] = - {\rm i} \hbar \{(1 - \beta P_{\rho} P^{\rho}) g^{\mu\nu} - \beta' 
  [P^{\mu} P^{\nu} + (\delta P^{\mu}) P^{\nu} + P^{\mu} \delta P^{\nu}]\}
\end{equation}
which can be rewritten as
\begin{equation}
  [X^{\prime\mu}, P^{\prime\nu}] = - {\rm i} \hbar [(1 - \beta P'_{\rho} P^{\prime\rho}) g^{\mu\nu}
  - \beta' P^{\prime\mu} P^{\prime\nu}]
\end{equation}
since $P'_{\rho} P^{\prime\rho} = P_{\rho} P^{\rho}$. Hence the first commutation relation in
(\ref{eq:alg-com}) is form invariant. A similar result can be easily proved for the remaining two commutation
relations.  This explicitly shows the invariance of our deformed algebra under proper Lorentz
transformations.\par
%
%
The generators $\hl_{\alpha\beta}$ of such transformations, satisfying the properties
\begin{equation}
  \delta X^{\mu} = \frac{\rm i}{2\hbar} \delta \omega^{\alpha\beta} \left[\hl_{\alpha\beta}, X^{\mu}
  \right] \qquad   \delta P^{\mu} = \frac{\rm i}{2\hbar} \delta \omega^{\alpha\beta}
  \left[\hl_{\alpha\beta}, P^{\mu}\right]  
\end{equation}
are given by
\begin{equation}
  \hl_{\alpha\beta} = (1 - \beta P_{\rho} P^{\rho})^{-1} (X_{\alpha} P_{\beta} - X_{\beta} P_{\alpha})
  = (1 - \beta P_{\rho} P^{\rho})^{-1} (P_{\beta} X_{\alpha} - P_{\alpha} X_{\beta}) 
  \label{eq:Lorentz-gen}
\end{equation}
and fulfil the standard so($D$,1) commutation relations
\begin{equation}
  \left[\hl_{\alpha\beta}, \hl_{\rho\sigma}\right] = - {\rm i} \hbar \left(g_{\alpha\rho} \hl_{\beta\sigma}
  - g_{\alpha\sigma} \hl_{\beta\rho} - g_{\beta\rho} \hl_{\alpha\sigma} + g_{\beta\sigma}
  \hl_{\alpha\rho}\right).
\end{equation}
\par
%
%
We conclude that, as far as proper Lorentz transformations are concerned, everything works as in the case
of conventional canonical commutation relations except for the generators, which get deformed. It is worth
observing that the deformed angular momentum operators $\hl_{ij}$ for the nonrelativistic algebra
(\ref{eq:Kempf-com}) \cite{kempf95} could be obtained from the corresponding operators
(\ref{eq:Lorentz-gen}) by dropping the term $(P^0)^2$ from $P_{\rho} P^{\rho}$, as already noted above
for the algebra commutation relations.\par
%
%
Our conclusion can be easily extended to improper Lorentz transformations since the discrete symmetries,
namely parity
\begin{equation}
  P: \qquad X^0 \to X^0 \qquad X^i \to - X^i \qquad P^0 \to P^0 \qquad P^i \to - P^i
\end{equation}
and time reversal
\begin{equation}
  T: \qquad X^0 \to - X^0 \qquad X^i \to X^i \qquad P^0 \to P^0 \qquad P^i \to - P^i \qquad {\rm i} \to -
  {\rm i} 
\end{equation}
obviously leave equation (\ref{eq:alg-com}) invariant.\par
%
%
Infinitesimal translations are more difficult to deal with because to ensure that the algebra commutation
relations remain invariant, one has to deform the standard transformations $x^{\prime\mu} = x^{\mu} -
\delta a^{\mu}$, $p^{\prime\mu} = p^{\mu}$ into
\begin{eqnarray}
  && X^{\prime\mu} = X^{\mu} + \delta X^{\mu} \qquad \delta X^{\mu} = - \delta a^{\mu} - g(P_{\rho}
         P^{\rho}) \delta a_{\nu} P^{\nu} P^{\mu} \nonumber \\
  && P^{\prime\mu} = P^{\mu} + \delta P^{\mu} \qquad \delta P^{\mu} = 0 \label{eq:transl-tr}  
\end{eqnarray}
with
\begin{equation}
  \delta a^{\mu} \in \mathbf{R} \qquad g(P_{\rho} P^{\rho}) = \frac{2\beta - \beta' - (2\beta + \beta') 
        \beta P_{\rho} P^{\rho}}{(1 - \beta P_{\rho} P^{\rho})^2}. 
\end{equation}
As a consequence, if we consider two particles with coordinates $x_1^{\mu}$ and $x_2^{\mu}$,
respectively, the difference $x_1^{\mu} - x_2^{\mu}$, which remains invariant under any conventional
translation, is transformed into an operator $X_1^{\mu} - X_2^{\mu}$, which, under some deformed
translation, will change and become dependent on the momentum operators $P_1^{\mu}$ and
$P_2^{\mu}$ of the two particles.\par
%
%
Transformation (\ref{eq:transl-tr}) results from the action of the generators
\begin{equation}
  \hp_{\alpha} = (1 - \beta P_{\rho} P^{\rho})^{-1} P_{\alpha}
\end{equation}
since $\delta X^{\mu}$ and $\delta P^{\mu}$ in (\ref{eq:transl-tr}) satisfy the relations
\begin{equation}
  \delta X^{\mu} = \frac{\rm i}{\hbar} \delta a^{\alpha} \left[\hp_{\alpha}, X^{\mu}\right] \qquad   
  \delta P^{\mu} = \frac{\rm i}{\hbar} \delta a^{\alpha} \left[\hp_{\alpha}, P^{\mu}\right].  
\end{equation} 
\par
%
%
As in the case of $\hl_{\alpha\beta}$, the commutation relations of the operators $\hp_{\alpha}$
\begin{equation}
  \left[\hp_{\alpha}, \hp_{\beta}\right] = 0
\end{equation}
are not disturbed by the deformation. The same is true for the mixed commutation relations
\begin{equation}
  \left[\hl_{\alpha\beta}, \hp_{\rho}\right] = {\rm i} \hbar \left(g_{\beta\rho} \hp_{\alpha} -
  g_{\alpha\rho} \hp_{\beta}\right). 
\end{equation}
The deformed operators $\hl_{\alpha\beta}$ and $\hp_{\alpha}$ therefore provide us with a realization of
the conventional Poincar\'e algebra iso($D$,1), leaving the commutation relations (\ref{eq:alg-com})
invariant. Observe that our results extend to the more general algebra (\ref{eq:alg-com}) those recently
obtained for Snyder algebra~\cite{banerjee}.\par
%
%
\subsection{Uncertainty relations}

Let us consider the uncertainty relation for position and momentum. Since the deformed algebra
(\ref{eq:alg-com}) is invariant under rotations, it is enough to consider a given pair of position and
momentum components, $X^i$, $P^i$, for some $i \in \{1, 2, \ldots, D\}$. We then get the inequality
\begin{equation}
  \Delta X^i \Delta P^i \ge \frac{\hbar}{2} \left|1 - \beta \left\{\langle (P^0)^2 \rangle - \sum_{j=1}^D
  \left[(\Delta P^j)^2 + \langle P^j \rangle^2\right] \right\} + \beta' \left[(\Delta P^i)^2 + \langle P^i
  \rangle^2\right]\right|.  \label{eq:UR}
\end{equation}
\par
%
%
On assuming isotropic uncertainties $\Delta P^j = \Delta P$, $j=1$, 2,~\ldots, $D$, for simplicity's sake,
equation (\ref{eq:UR}) yields
\begin{equation}
  \Delta X^i \ge \frac{\hbar}{2} \left| \frac{1 - \beta \left[\langle (P^0)^2 \rangle - \sum_{j=1}^D
  \langle P^j \rangle^2\right] + \beta' \langle P^i \rangle^2}{\Delta P} + (D\beta + \beta') \Delta P\right|. 
\end{equation}
From this, it follows that $\Delta X^i$ has a nonvanishing minimum
\begin{equation}
  \Delta X^i_{\rm min} = \hbar \sqrt{(D\beta + \beta') \left\{1 - \beta \left[\langle (P^0)^2 \rangle -
  \sum_{j=1}^D \langle P^j \rangle^2\right] + \beta' \langle P^i \rangle^2 \right\}}
\end{equation}
provided the quantity between curly brackets on the right-hand side is positive. This condition is always
satisfied by physically acceptable states due to relation (\ref{eq:cond-accept}). We therefore arrive at an
isotropic absolutely smallest uncertainty in position given by
\begin{equation}
  (\Delta X)_0 = (\Delta X^i)_0 = \hbar \sqrt{(D\beta + \beta') \left[1 - \beta \langle (P^0)^2 \rangle
  \right]}. 
\end{equation}
As compared with Kempf's result~\cite{kempf97}, there is an additional factor $\sqrt{1 - \beta \langle
(P^0)^2 \rangle}$ reducing $(\Delta X)_0$.\par
%
%
We might also formally use the Heisenberg uncertainty relation for time and energy to look into the possible
existence of a nonzero minimal uncertainty in time. We shall, however, refrain from doing so, because
time-energy uncertainty has a particular status in quantum mechanics. It does not follow from the
commutation relation of two operators (see, e.g., \cite{aharonov}), but requires a more careful and detailed
investigation, which we think is worth separate papers.\par
%
%
\section{\boldmath ($1+1$)-dimensional Dirac oscillator with Lorentz-covariant deformed algebra}
\setcounter{equation}{0}

Since the deformed algebra (\ref{eq:alg-com}) remains invariant under the standard Lorentz transformations
(\ref{eq:Lorentz-tr}), (\ref{eq:omega}), it is clear that substituting $X^{\mu}$, $P^{\mu}$ for $x^{\mu}$,
$p^{\mu}$ in the conventional Dirac equation will provide us with a Lorentz-covariant extension of this
equation. In particular, the proof of the Dirac oscillator Lorentz covariance, given in~\cite{moreno}, remains
true in our generalized context. In the present section, we plan to study such a system in the simplest case,
corresponding to a ($1+1$)-dimensional quantized spacetime when one of the deforming parameters
vanishes and the other one is positive, namely $\beta'=0$ and $\beta>0$. Under such conditions, there is a
nonzero minimal uncertainty in position.\par
%
%
In momentum representation, the equation to be solved reads
\begin{equation}
  \left[\hat{\alpha}_x \left(P - {\rm i} m\omega \hat{\beta} X\right) + mc \hat{\beta}\right] \psi(p,
  p^0) = P^0 \psi(p, p^0)  \label{eq:DO}
\end{equation}
where $X = X^1$, $P = P^1$ and $P^0$ are given by (\ref{eq:alg-gen}) and satisfy the commutation 
relations (\ref{eq:alg-com}) for $D=1$, $\beta'=0$ and $\beta > 0$. For the $2 \times 2$ matrices
$\hat{\alpha}_x$ and $\hat{\beta}$, we can take $\hat{\alpha}_x = \sigma_x$ and $\hat{\beta} =
\sigma_z$, where $\sigma_x$ and $\sigma_z$ denote standard Pauli spin matrices. Hence, equation
(\ref{eq:DO}) can be rewritten as
\begin{equation}
  (\sigma_x P - m\omega \sigma_y X + mc \sigma_z) \psi(p, p^0) = P^0 \psi(p, p^0).  \label{eq:DO-bis} 
\end{equation}
\par
%
%
It is convenient to introduce dimensionless position and momentum operators
\begin{equation}
  \txx^{\mu} = \frac{X^{\mu}}{a} \qquad \tpp^{\mu} = \frac{a}{\hbar} P^{\mu} \qquad \mu = 0, 1
\end{equation}
where the length unit is $a = \hbar/(mc)$. Such operators satisfy the commutation relations
\begin{eqnarray}
  && \left[\txx^0, \tpp^0\right] = - \left[\txx, \tpp\right] = - {\rm i}\left\{1 - \tbeta \left[\left(\tpp^0
        \right)^2 - \tpp^2\right]\right\} \nonumber \\
  && \left[\txx^0, \txx\right] = 2{\rm i} \tbeta \left(\tpp^0 \txx - \tpp \txx^0\right) \nonumber \\  
  && \left[\txx^0, \tpp\right] =  \left[\txx, \tpp^0\right] =  \left[\tpp^0, \tpp\right] = 0
        \label{eq:alg-com-1}  
\end{eqnarray}
where we have set $\tbeta = \beta m^2 c^2$. Equation (\ref{eq:DO-bis}) now becomes
\begin{equation}
  (\sigma_x \tpp - \tomega \sigma_y \txx + \sigma_z) \psi(\tp, \tp^0) = \tpp^0 \psi(\tp, \tp^0) 
  \label{eq:DO-ter} 
\end{equation}
where $\tp^{\mu} = (a/\hbar) p^{\mu}$, $\tomega = \hbar\omega/(mc^2)$ and
\begin{eqnarray}
  && \tpp^0 = \tp^0 \qquad \tpp = \tp  \nonumber \\
  && \txx^0 = - {\rm i} f(\tp, \tp^0) \frac{\partial}{\partial \tp^0} \qquad \txx = {\rm i} f(\tp, \tp^0)
       \frac{\partial}{\partial \tp} \qquad f(\tp, \tp^0) = 1 - \tbeta [(\tp^0)^2 - \tp^2] 
\end{eqnarray}
with $\gamma \equiv 0$.\par
%
%
On separating the wavefunction $\psi = \left(\begin{array}{c} \psi_1 \\ \psi_2 \end{array}\right)$ in
(\ref{eq:DO-ter}) into large $\psi_1$ and small $\psi_2$ components, the Dirac oscillator equation can be
written as two coupled equations
\begin{eqnarray}
  && B^+ \psi_2(\tp, \tp^0) = (\tp^0 - 1) \psi_1(\tp, \tp^0) \label{eq:DO-1}\\
  && B^- \psi_1(\tp, \tp^0) = (\tp^0 + 1) \psi_2(\tp, \tp^0)  \label{eq:DO-2}
\end{eqnarray}
where, as a consequence of the $\beta'=0$ assumption,
\begin{equation}
  B^{\pm} = \tpp \pm {\rm i} \tomega \txx = \tp \mp \tomega f(\tp, \tp^0) \frac{\partial}{\partial \tp}
\end{equation}
are differential operators in $\tp$ only (but containing a dependence on $\tp^0$).\par
%
%
\subsection{Energy spectrum}

Let us determine for which values of $\tp^0$ equations (\ref{eq:DO-1}) and (\ref{eq:DO-2}) are
compatible. For such a purpose, we note that $B^{\pm}$ do not act on $\tp^0 \pm 1$. Hence, on
applying $B^+$ (respectively $B^-$) to (\ref{eq:DO-2}) (respectively (\ref{eq:DO-1})) and using
(\ref{eq:DO-1}) (respectively (\ref{eq:DO-2})), we get the following factorized equation for the large
component $\psi_1$ (respectively small component $\psi_2$)
\begin{eqnarray}
  && B^+ B^- \psi_1(\tp, \tp^0) = e(\tp^0) \psi_1(\tp, \tp^0)  \label{eq:SUSY-1} \\
  && B^- B^+ \psi_2(\tp, \tp^0) = e(\tp^0) \psi_2(\tp, \tp^0)  \label{eq:SUSY-2}
\end{eqnarray}
where
\begin{equation}
  e(\tp^0) = (\tp^0)^2 - 1.  \label{eq:e}
\end{equation}
\par
%
%
These equations have the form of energy-eigenvalue equations in SUSYQM~\cite{cooper, junker}. It should be
stressed, however, that the starting Hamiltonian $H = B^+ B^-$ depends on $(\tp^0)^2$, which determines
the eigenvalues $e(\tp^0)$. Hence, in contrast with conventional SUSYQM, $H$ is energy dependent. As we
plan to show, the shape-invariance method, originally developed in the usual framework and generalized to
deformed algebras with minimal length in \cite{cq03, cq04, cq05}, can be applied to energy-dependent
Hamiltonians.\par
%
%
{}For such a purpose, let us consider a hierarchy of Hamiltonians
\begin{equation}
  H_i = B^+(g_i) B^-(g_i) + \sum_{j=0}^i \epsilon_j \qquad i = 0, 1, 2, \ldots  \label{eq:B_i}  
\end{equation}
whose first member $H_0$ coincides with $H$. Here
\begin{equation}
  B^{\pm}(g_i) = g_i \tpp \pm {\rm i} \tomega \txx = g_i \tp \mp \tomega f(\tp, \tp^0)
  \frac{\partial}{\partial \tp}  \label{eq:B}
\end{equation}
$g_i$, $\epsilon_i$, $i=0$, 1, 2,~\ldots, are real and $g_0=1$, $\epsilon_0=0$. The quantities $g_i$,
$\epsilon_i$, $i=1$, 2,~\ldots, will be determined through a shape-invariance condition
\begin{equation}
  B^-(g_i) B^+(g_i) = B^+(g_{i+1}) B^-(g_{i+1}) + \epsilon_{i+1}  \qquad i = 0, 1, 2, \ldots.
  \label{eq:SI}
\end{equation}
On inserting (\ref{eq:B}) in (\ref{eq:SI}), we directly get the set of two relations
\begin{eqnarray}
  && g_{i+1} (g_{i+1} - \tbeta \tomega) = g_i (g_i + \tbeta \tomega)  \label{eq:C1} \\
  && \epsilon_{i+1} = \tomega (g_i + g_{i+1}) [1 - \tbeta (\tp^0)^2].
\end{eqnarray}
\par
%
%
Such equations are easily solved. Among the two solutions of (\ref{eq:C1}), $g_{i+1} = g_i + \tbeta
\tomega$ and $g_{i+1} = - g_i$, we choose the former because, in section~3.2, it will be shown that the
wavefunction normalizability imposes the condition $g_i > 0$. We therefore obtain
\begin{eqnarray}
  && g_i = 1 + \tbeta \tomega i \label{eq:g}\\
  && \epsilon_{i+1}(\tp^0) = \tomega [2 + \tbeta\tomega (2i+1)][1 - \tbeta (\tp^0)^2]
       \label{eq:epsilon}
\end{eqnarray}
for $i=0$, 1, 2,~\ldots. Observe that in contrast with what happens in conventional SUSYQM,
$\epsilon_{i+1}$ is not a constant, but a function of $\tp^0$.\par
%
%
{}From this, we find that the eigenvalues of the SUSYQM Hamiltonian $H = B^+ B^-$ in (\ref{eq:SUSY-1}) are
given by
\begin{equation}
  e_n(\tp^0) = \sum_{i=0}^n \epsilon_i(\tp^0) = \tomega n (2 + \tbeta\tomega n) [1 - \tbeta (\tp^0)^2].
  \label{eq:e-bis}
\end{equation}
For such a result to be compatible with the definition of $e(\tp^0)$, given in (\ref{eq:e}), $\tp^0$ must be
quantized and its allowed values given by
\begin{equation}
  \tp^0_{n, \tau} = \tau \left(\frac{1 + \tomega n (2 + \tbeta\tomega n)}{1 + \tbeta\tomega n (2 +
  \tbeta\tomega n)}\right)^{1/2} = \frac{\tau}{\sqrt{\tbeta}} \left(1 + \frac{\tbeta - 1}{(1 +
  \tbeta\tomega n)^2}\right)^{1/2}  \label{eq:p^0} 
\end{equation}
where $\tau = \pm 1$ and $n$ may, in principle, run over $n=0$, 1, 2,~\ldots. On substituting this
expression for $\tp^0$ in (\ref{eq:e}) or (\ref{eq:e-bis}), the energy spectrum of $H$ acquires the form
\begin{equation}
  e_n \equiv e_n(\tp^0_{n, \tau}) = \frac{(1 - \tbeta) \tomega n (2 + \tbeta\tomega n)}{(1 +
  \tbeta\tomega n)^2} = \frac{1 - \tbeta}{\tbeta} \left(1 - \frac{1}{(1 + \tbeta\tomega n)^2}\right).
\end{equation}
\par
%
%
On the other hand, since $E = c p^0 = m c^2 \tp^0$, the Dirac oscillator energy spectrum is given by
\begin{equation}
  E_{n, \tau} = \tau mc^2 \left(\frac{1 + \hbar\omega n (2 + \beta m\hbar\omega n)/(mc^2)}{1 +
  \beta m\hbar\omega n (2 + \beta m\hbar\omega n)}\right)^{1/2} = \frac{\tau c}{\sqrt{\beta}} \left(1 +
  \frac{\beta m^2 c^2 - 1}{(1 + \beta m\hbar\omega n)^2}\right)^{1/2}  \label{eq:E}
\end{equation}
where the actual values of $(n, \tau)$ depend on the existence of normalizable solutions to equations
(\ref{eq:DO-1}) and (\ref{eq:DO-2}). In section~3.2, it will be proved that $n=0$, 1, 2,~\ldots, or $n=1$,
2,~\ldots, according to whether $\tau = +1$ or $\tau = -1$.\par
%
%
{}From equation (\ref{eq:E}), we see that if $\beta$ could take values such that $\beta > 1/(m^2c^2)$,
the energy spectrum would have an unphysical behaviour in the sense that $|E_{n,\tau}|$ would
decrease with increasing $n$. This indicates that the deforming parameter $\beta$ must satisfy the condition
$\beta < 1/(m^2c^2)$, in which case $|E_{n,\tau}|$ will increase with $n$. Note that this will be
confirmed in section~3.2 by the existence of well-behaved wavefunctions in such a range of $\beta$
values.\par
%
%
{}Furthermore, we observe that in contrast with the conventional one-dimensional Dirac
oscillator~\cite{toyama}, the energy spectrum is bounded:
\begin{equation}
  mc^2 \le |E_{n,\tau}| < \frac{c}{\sqrt{\beta}}
\end{equation}
the upper limit being attained for $n \to \infty$. Since $E_{n,\tau}^2 < c^2/\beta$ is equivalent to $\beta
(p^0_{n,\tau})^2 < 1$, this property actually agrees with equation (\ref{eq:cond-accept}) to be satisfied
by physically acceptable states. In the case where $\beta m\hbar\omega \ll 1$, equation (\ref{eq:E}) yields
\begin{equation}
  E_{n,\tau} \simeq \tau mc^2 \left(1 + \frac{2\hbar\omega}{mc^2} n\right)^{1/2} \left(1 - \frac{3}{2} 
  \beta \frac{(\hbar\omega n/c)^2}{1 + \frac{2\hbar\omega}{mc^2} n} + \cdots\right)
\end{equation}
so that when the deformation vanishes, we get back the unbounded energy spectrum of the conventional
Dirac oscillator~\cite{toyama}.\par
%
%
{}Finally, in the nonrelativistic limit $\hbar\omega/(mc^2) \ll 1$, equation (\ref{eq:E}) leads to
\begin{equation}
  E_{n,\tau} \simeq \tau (1 + \beta m\hbar\omega n)^{-1} \left[mc^2 + \hbar\omega n \left(1 + 
  \frac{1}{2} \beta m\hbar\omega n\right) + \cdots \right].  \label{eq:nonrel-E} 
\end{equation}
One may observe that the quantity between square brackets in (\ref{eq:nonrel-E}) corresponds to the
nonrelativistic limit of the Dirac oscillator energy spectrum that would be obtained using the
non-Lorentz-covariant Kempf algebra along the same lines as in~\cite{cq05}. The presence of the additional
factor $(1 + \beta m\hbar\omega n)^{-1}$ in (\ref{eq:nonrel-E}) is another evidence of the novelty of the
Lorentz-covariant deformed algebra proposed in this paper, as compared with Kempf one (see discussion
below equation (\ref{eq:alg-com}).\par
%
%
\subsection{Wavefunctions}

In this subsection, we will begin by employing SUSYQM methods to calculate the wavefunctions of the
Hamiltonians $H_i$, defined in (\ref{eq:B_i}). At the end, we will go back to the determination of large and
small components of the Dirac oscillator wavefunction, satisfying (\ref{eq:DO-1}) and (\ref{eq:DO-2}).\par
%
%
{}For some given quantized value $(\tp^0_{n,\tau})^2$ of $(\tp^0)^2$, let us denote the operators
(\ref{eq:B}) and the functions (\ref{eq:epsilon}) by $B^{\pm}_n(g_i)$ and $\epsilon_{i+1,n}$,
respectively\footnote{It should be stressed that the operators and functions considered in the SUSYQM
approach do not depend on $\tau$. Such a dependence will only appear when going back to the Dirac
oscillator equation, where $\tau$ is interpreted as the energy sign.}. Then, in the case where SUSYQM is
unbroken and $B^-_0(g) \phi^{(0)}(g;\tp) = 0$ has a normalizable solution, the eigenvalue problem for $H$
and, more generally, for $H_i$ takes the form
\begin{eqnarray}
  && H \phi^{(n)}(g;\tp) = e_n \phi^{(n)}(g;\tp) \nonumber \\
  && H_i \phi^{(n-i)}(g_i;\tp) = e_n \phi^{(n-i)}(g_i;\tp)  \label{eq:SUSY-bis}
\end{eqnarray}
where, for $H$ and $H_i$, we substitute $B^+_n(g) B^-_n(g)$ and $B^+_n(g_i) B^-_n(g_i) +
\sum_{j=0}^i \epsilon_{j,n}$, respectively.\par
%
%
Let us first check that SUSYQM is unbroken by considering 
\begin{equation}
  B^-_i(g_i) \phi^{(0)}(g_i;\tp) = 0.
\end{equation}
Up to some multiplicative constant, the solution of this first-order differential equation can be written as
\begin{equation}
  \phi^{(0)}(g_i;\tp) = [f_i(\tp)]^{-g_i/(2\tbeta\tomega)}
\end{equation}
where we have set $f_i(\tp) = f(\tp,\tp^0_{i,\tau})$. Such a function is normalizable with respect to the
scalar product (\ref{eq:sc}) (with $\tp^0 \to \tp^0_{i,\tau}$) provided $g_i > 0$, which corresponds to
the choice made in (\ref{eq:g}). In addition, it can be easily shown that for $g_i > 0$, $B^+(g_i)$ has no
normalizable zero mode.\par
%
%
In particular, for $i=0$, we get for the ground-state wavefunction of $H$
\begin{equation}
  \phi^{(0)}(g;\tp) = [f_0(\tp)]^{-g/(2\tbeta\tomega)}  \label{eq:phi-0}
\end{equation}
where $g=1$ and $f_0(\tp) = 1 - \tbeta + \tbeta \tp^2$. It is clear that $\phi^{(0)}(g;\tp)$ belongs to
the set of physical states since it is nonsingular for $\tbeta < 1$ or, equivalently, $\beta <
1/(m^2c^2)$.\par
%
%
We can now determine the excited-state wavefunctions of $H$ by solving the recursion relation
\begin{equation}
  \phi^{(n)}(g;\tp) \propto B^+_n(g) \phi^{(n-1)}(g_1;\tp).  \label{eq:recursion}
\end{equation}
On using the ansatz
\begin{equation}
  \phi^{(n)}(g;\tp) \propto [f_n(\tp)]^{-g/(2\tbeta\tomega)} P_n(g;z)  \label{eq:phi}
\end{equation}
where $P_n(g;z)$ is some function of $z = \tp \sqrt{\tbeta/f_n(\tp)}$, such that $P_0(g_n;z) = 1$, we
obtain that equation (\ref{eq:recursion}) is equivalent to
\begin{equation}
  P_n(g;z) \propto \left[(1 - z^2) \frac{d}{dz} + \left(1 - 2 \frac{g_1}{\tbeta\tomega}\right)z\right]
  P_{n-1}(g_1;z).
\end{equation}
Hence, $P_n(g;z)$ is some $n$th-degree polynomial in $z$. Furthermore, comparison with equation
(1.8.21) of~\cite{koekoek} allows us to identify it with a Gegenbauer polynomial:
\begin{equation}
  P_n(g;z) = C^{(\lambda)}_n(z) \qquad \lambda = \frac{g}{\tbeta\tomega} = \frac{1}{\tbeta\tomega}.
  \label{eq:P}
\end{equation}
Here we have used the fact that replacing $g$ by $g_1 = g + \tbeta\tomega$ and $n$ by $n-1$ in
(\ref{eq:P}) yields $P_{n-1}(g_1;z) = C^{(\lambda_1)}_{n-1}(z)$ with $\lambda_1 = \lambda + 1$.\par
%
%
On collecting the results contained in (\ref{eq:SUSY-bis}), (\ref{eq:phi}) and (\ref{eq:P}), we can write the
normalizable wavefunctions of $H$ and of its SUSYQM partner $H_1$, corresponding to the same energy
$e_n$, as
\begin{eqnarray}
  && \phi^{(n)}(g;\tp) = [f_n(\tp)]^{-\lambda/2} C^{(\lambda)}_n(z) \label{eq:phi-1} \\
  && \phi^{(n-1)}(g_1;\tp) = [f_n(\tp)]^{-(\lambda+1)/2} C^{(\lambda+1)}_{n-1}(z)  \label{eq:phi-2}
\end{eqnarray}
where, in (\ref{eq:phi-1}), $n$ runs over $n=0$, 1, 2,~\ldots, whereas, in (\ref{eq:phi-2}), it is restricted to
the values $n=1$, 2,~\ldots.\par
%
%
Let us now go back to the Dirac oscillator problem. From equations (\ref{eq:SUSY-1}), (\ref{eq:SUSY-2}) and
(\ref{eq:SI}), it is expected that the large and small components of the Dirac oscillator wavefunctions,
associated with a given $e_n$, be given by equations (\ref{eq:phi-1}) and (\ref{eq:phi-2}), respectively.
However, one should keep in mind that though the solutions of equations (\ref{eq:DO-1}) and
(\ref{eq:DO-2}) provide us with solutions to equations (\ref{eq:SUSY-1}) and (\ref{eq:SUSY-2}), the
converse is not necessarily true. Furthermore, it is clear from (\ref{eq:DO-1}) and (\ref{eq:DO-2}) that
whereas the solutions of (\ref{eq:SUSY-1}) and (\ref{eq:SUSY-2}) are the same for positive and negative
values of $\tp^0$, the large and small components will have a dependence on the sign of $\tp^0$. Hence,
we shall denote them as $\psi_1^{(n,\tau)}(\tp) \equiv \psi_1(\tp,\tp^0_{n,\tau})$ and
$\psi_2^{(n,\tau)}(\tp) \equiv \psi_2(\tp,\tp^0_{n,\tau})$.\par
%
%
Let us start with $\phi^{(0)}(g;\tp)$, for which $(\tp^0)^2 = 1$ and which should correspond to the large
component of the Dirac oscillator ground-state wavefunction. Since $B^+(g)$ has no normalizable zero
mode, it is clear that equations (\ref{eq:DO-1}) and (\ref{eq:DO-2}) are fulfilled by
\begin{equation}
  \psi_1^{(0,+1)}(\tp) = N_1^{(0,+1)} \phi^{(0)}(g;\tp) \qquad \psi_2^{(0,+1)}(\tp) = 0
  \label{eq:DO-gs}
\end{equation}
for $\tp^0_{0,+1} = +1$, but that they have no solution for $\tp^0_{0,-1} = -1$. In (\ref{eq:DO-gs}),
$N_1^{(0,+1)}$ denotes some normalization constant.\par
%
%
{}From known properties of Gegenbauer polynomials~\cite{koekoek}, it is then straightforward to check that
the functions
\begin{equation}
  \psi_1^{(n,\tau)}(\tp) = N_1^{(n,\tau)} \phi^{(n)}(g;\tp) \qquad \psi_2^{(n,\tau)}(\tp) =
  N_2^{(n,\tau)} \phi^{(n-1)}(g_1;\tp)
  \label{eq:DO-es}
\end{equation}
satisfy equations (\ref{eq:DO-1}) and (\ref{eq:DO-2}) for $n=1$, 2,~\ldots, and $\tau = \pm 1$, provided
their normalization factors are connected by the relation
\begin{equation}
  N_2^{(n,\tau)} = \frac{2\left[1 - \tbeta (\tp^0_{n,\tau})^2\right]}{\sqrt{\tbeta}\, (\tp^0_{n,\tau} + 1)}
  N_1^{(n,\tau)}.  \label{eq:N-N}
\end{equation}
\par
%
%
The yet undetermined constants $N_1^{(n,\tau)}$, $(n,\tau) = (0, +1), (1, \pm 1), (2, \pm 1)$,~\ldots,
can be calculated from the normalization condition
\begin{equation}
  \int_{-\infty}^{+\infty} \frac{d\tp}{f_n(\tp)} \left[\left|\psi_1^{(n,\tau)}(\tp)\right|^2 +
  \left|\psi_2^{(n,\tau)}(\tp)\right|^2\right] = 1.
\end{equation}
The result reads
\begin{eqnarray}
  && N_1^{(n,\tau)} = \left(\frac{\tp^0_{n,\tau} + 1}{2\tp^0_{n,\tau}}\right)^{1/2} A^{(n)}(\lambda)
        \nonumber \\
  && A^{(n)}(\lambda) = 2^{\lambda} \Gamma(\lambda) \left(\frac{\sqrt{\tbeta}\, (\lambda + n) n!
        \left[1 - \tbeta \left(\tp^0_{n,\tau}\right)^2\right]^{\lambda + \frac{1}{2}}}{2\pi \Gamma(2\lambda
        +n)}\right)^{1/2}  \label{eq:A}
\end{eqnarray}
with $1 - \tbeta \left(\tp^0_{n,\tau}\right)^2 = (1 - \tbeta) \lambda^2/(\lambda + n)^2$ and
$\tp^0_{n,\tau}$ given in (\ref{eq:p^0}). On combining (\ref{eq:N-N}) with (\ref{eq:A}), we also get
\begin{equation}
  N_2^{(n,\tau)} = \tau \left(\frac{\tp^0_{n,\tau} - 1}{2\tp^0_{n,\tau}}\right)^{1/2}
  A^{(n-1)}(\lambda+1).  \label{eq:N_2} 
\end{equation}
\par
%
%
Equations (\ref{eq:phi-0}), (\ref{eq:phi-1})--(\ref{eq:DO-es}), (\ref{eq:A}) and (\ref{eq:N_2}) therefore
yield the whole set of positive- and negative-energy normalized wavefunctions of the Dirac oscillator in the
($1+1$)-dimensional quantized spacetime defined by equation (\ref{eq:alg-com-1}). As previously shown,
such bound-state wavefunctions exist in the parameter range $\beta < 1/(m^2c^2)$. As a final point, we
would like to emphasize that due to the energy dependence of the Hamiltonian, the standard orthogonality
relation between separate bound states is lost. This problem is one of the many known puzzles inherent in
the use of energy-dependent Hamiltonians in both nonrelativistic and relativistic quantum mechanics (for
some recent reviews see, e.g., ~\cite{formanek, znojil}). In the latter context, it appeared along with the
Klein-Gordon~\cite{snyder40} and Bethe-Salpeter~\cite{bethe} equations many years ago and it has been
the topic of several investigations (see, e.g.,~\cite{sazdjian}). In view of its complexity, we will leave it for
future studies.\par
%
%
\section{Conclusion}

In this paper, we have generalized the $D$-dimensional $(\beta, \beta')$-two-parameter deformed algebra
with minimal length introduced by Kempf to a Lorentz-covariant algebra in a ($D+1$)-dimensional quantized
spacetime. In the $D=3$ and $\beta=0$ case, it reproduces Snyder algebra.\par
%
%
We have obtained the deformed Poincar\'e transformations leaving our algebra invariant and proved that, in
contrast with Lorentz transformations, translations get deformed. Although the Poincar\'e generators
acquire a $\beta$ dependence, they close the usual Poincar\'e algebra. The Lorentz generators generalize
Kempf deformed angular momentum operators, while, in the special case of Snyder algebra, all Poincar\'e
generators retain their undeformed structure, thereby confirming a recent, independent result.\par
%
%
Our study of uncertainty relations has shown that there exist (isotropic) nonzero minimal uncertainties in the
position coordinates (minimal length). As compared with Kempf's result, they contain an additional
energy-dependent reducing factor.\par
%
%
We have considered one of the simplest relativistic quantum systems, namely the Dirac oscillator in a
($1+1$)-dimensional quantized spacetime, described by our Lorentz-covariant deformed algebra with minimal
length, in the case where $\beta > 0$ and $\beta'=0$. Extending standard SUSYQM and shape-invariance
methods to energy-dependent Hamiltonians has provided us with the exact solutions to the
energy-eigenvalue problem for such a system. We have shown that physically acceptable bound states exist
for $\beta < 1/(m^2c^2)$. A new interesting outcome arising from our study has been the energy-spectrum
boundedness: in contrast with the conventional Dirac oscillator, the energy absolute value $|E_{n\tau}|$
indeed goes to a finite limit $c/\sqrt{\beta}$ for $n \to \infty$.\par
%
%
In passing we have pointed out two remaining problems for future investigations: on one hand, a thorough
analysis of the time-energy uncertainty, required by its special status in quantum mechanics,
and, one the other hand, an attempt to restore some of the properties of ordinary quantum mechanics that
are spoilt by the energy dependence of the Dirac oscillator Hamiltonian, such as the orthogonality of bound
states corresponding to different energies.\par
%
%
As a final point, we would like to signal the relevance of interpreting the physical significance
of the two deforming parameters $\beta$, $\beta'$, appearing in our algebra. As we already observed,
$\beta'$ is connected with Snyder deforming parameter. Furthermore, it has been shown~\cite{kowalski}
that the latter can be related through the quantum $\kappa$-Poincar\'e algebra~\cite{lukierski} to the
second observer-independent scale predicted (along with $c$) by doubly special relativity~\cite{amelino}. An
interesting point would therefore be the search for a physical interpretation of the remaining parameter
$\beta$. Among other things, this might provide us with some argument to implement condition
(\ref{eq:cond-accept}) on physically acceptable states, which would be stronger than those given in
section~3.\par
%
%
\section*{Acknowledgments}

CQ is a Research Director of the National Fund for Scientific Research (FNRS), Belgium.
VMT thanks this organization for financial support.\par
%
%

\end{document}